\documentstyle[11pt,paspconf,epsf]{article}

\begin{document}

\title{DETECTING PLANETS THROUGH MICROLENSING}
\smallskip
{\bf{\centerline{(To be published in the proceedings of}}}
{\it{\centerline{Planets Beyond the Solar System }}}
{\it{\centerline{and the Next Generation of Space Missions}}}
{\bf{\centerline{held in October 1996, at STScI, Baltimore, MD 21218)}}}
\vskip 1.3cm
\author{Kailash C. Sahu}
\affil{Space Telescope Science Institute, 3700 San Martin Drive, 
Baltimore, MD 21218}
\vskip 0.3cm

\begin{abstract}

More than 100 microlensing events have been detected during the
last ~4 years, most of them towards the Galactic Bulge. Since the
line of sight towards the Bulge passes through the disk and the 
Bulge itself, the known stars towards the Bulge play a dominant 
role as gravitational lenses. If these stars have planets around 
them, then the signature of the planets can be seen as sharp, 
extra peaks on the microlensing light curves. Frequent, 
continuous monitoring of the on-going microlensing events thus 
provides a powerful new method to search for planets around 
lensing stars.

Here I  first review the background on stars acting as 
gravitational lenses. I then review the theoretical work on
possible observational features due to planets, and the 
probability of detecting the planets through microlensing. 
I then discuss the status/strategy/results of the observational 
programs currently active in this field. 

\end{abstract}


\section{INTRODUCTION}

If we consider all the properties of the planets which have 
contributed to the discoveries of planets during the last
two hundred years, both within the solar system and outside, 
gravity is a clear winner. Whether it is the discovery 
of Neptune
about 150 years ago, or the discovery of Pluto about 65 years ago, 
or the discovery of planets around the pulsar PSR1257+12 about 
3 years ago, or the most recent discovery of planets around many 
nearby stars (See Review by Latham, this volume),
there is one common denominator in 
all these discoveries: all these have been discovered due to the 
gravitational effect of the planet. 

In 1845, the French astronomer Leverrier predicted the position
of Neptune from the orbital perturbations of Uranus. 
The prediction was then observationally followed up by Johan Galle, who
discovered Neptune in a single night of observations.
A similar story was repeated in 1930, when Lowell predicted
the position of Pluto from the orbital perturbations of Neptune, which
was then easily discovered by Tombaugh. The discovery of the first 
definitive extra-solar planet around the pulsar PSR1257+12 was again 
through the gravitational effect of the planet (Wolszczan and Frail, 1992) 
The most recent flurry of discoveries of planets around 
the nearby stars have made use of the gravitational effect 
in a different facet, namely the radial velocity 
perturbation it causes on the parent star (Mayor et al. 1995;
Mercey and Butler, 1995; Latham, this volume).
On the other hand, a tremendous amount of
effort has been spent in looking for planets around 
other stars though other esoteric means, such as spatial
interferometry or adaptive optics. 
While some of these efforts will no doubt bear fruit in the 
near future as we overcome the technical challenges they pose, 
they have borne very little fruit so far.
The reason is not difficult to understand: the gravitational effect,
in almost all cases, makes use of the bright nearby object whereas
the other methods seek to overcome the effect of the bright nearby
object through technology. In the case of spatial
interferometry or adaptive optics, one must always fight
to keep the light of the bright star down in order to
detect the faint planetary signal in the presence of this
highly dominant bright source. In other words, the bright star
always acts as a noise, is a hindrance to the search, 
and is always something that one must win over in order to
be able to detect the much fainter planet nearby.
The situation is reversed in case of the gravitational effect of the
planet, in which case, one simply uses the features in the brighter
object to look for perturbations.
In case of Neptune and Pluto, the nearby brighter object was used 
to look for perturbations in its orbit. In case of the pulsar PSR1257+12,
the pulse period distribution of the pulsar itself was used to 
look for the effect due to the planet.
And in case of the radial velocity measurements, 
the absorption lines from the parent star was necessary to
look for the effect of the planet.  

This paper discusses another aspect of the gravitational effect,
namely gravitational microlensing, the effect of which is similar
in the sense that, this too uses the brighter object nearby,  
the star in this case too, helps in the search 
for the planet nearby.  This may potentially be a very
powerful tool to look for extra-solar planets, and 
as discussed in
more detail later, this is the only method sensitive to the search 
for Earth-like planets around normal stars, using ground based 
observations.  
Furthermore, this is the only method which can provide a 
statistics on the masses and orbital radii of extra-solar planets.
It must be noted however that microlensing does have its
selection effects, and this method is more
sensitive to detection of planets around {\it low} mass stars
since, statistically, a large fraction of the lenses are expected to be
low mass stars.

The paper is structured as follows:
In section 2 and 3 of the paper I describe the details of the microlensing,
and the role of stars acting as lenses. In section 4, the basic
theoretical aspects of stars as lenses are briefly outlined, and the
effect of extended sources are described.
In section 5, the role of
planets as potential lenses are discussed, which  is then followed by
the details of the characteristic features due to
planets, the requirements for an observational program,
and the probability of detection in different search strategies.
Finally, the current observational programs 
towards this end are described, and some preliminary results
of the PLANET collaboration are presented.

\section{MICROLENSING}   

The idea of microlensing by stars is not new. In 1936, Einstein wrote a
small paper in {\it Science} 
where, he did `a little calculation' at the request of his friend
Mandal and showed that if a star happens to pass very close to another 
star in the line of sight, then the background star will be lensed
(Einstein, 1936). However, he also 
dismissed the idea as only a theoretical exercise and remarked that 
there was `no hope of observing such a phenomenon directly'. He was 
right at that time; the probability of
observing is  less than one in a million, and with the 
technology of 1936, there was no way one could observe this directly.

Paczy\'nski, in two papers written in 1986 and 1991, noted that if one 
could monitor a 
few million stars, one could observe microlensing events, perhaps
as a signature of the dark matter towards the LMC, or by known stars 
towards the Galactic Bulge (Paczy\'nski, 1986; Paczy\'nski, 1991). 
The project was taken up immediately by 
three groups and the first observed microlensing event was reported 
towards the LMC in 1993. By now, more than 100 events have been
discovered, mostly towards the Galactic Bulge.

\section{STARS AS LENSES}

\subsection{Towards the LMC}

After the first microlensing event was discovered towards the LMC
(Alcock et al. 1993; Aobourg et al. 1993), there
was great hope 
in the astronomical community that the illusive long-sought dark matter 
was finally found.
It was soon realised, however, that the observed optical depth to 
microlensing towards the LMC is too small for the dark matter in the halo 
to be made up of MACHOs. Microlensing by known stellar populations
towards the LMC were explored, and taking into account the 
number and distribution of the small number of the
observed events, it was argued that most of the lenses are probably
stars within the LMC itself (Sahu, 1994a; Sahu, 1994b; Wu, 1994).
It was also argued that some fraction of the events could be
due to stars within the local disk of our own Galaxy (Bahcall et al. 1994,
Gould et al, 1996, Flynn et al, 1996).
The number of events detected towards the LMC has grown, albeit slowly,
and a total of 8 events have been detected so far.
To date, no consensus has been reached on the exact location of the
lenses, nor on the contribution of the MACHOs to the dark halo, the claims
ranging from  0 to 50\%. All the 8 detected events come from the
MACHO group, who claim that the MACHO contribution to dark matter is 50\%
(Alcock et al., 1996), and the nature of the MACHOs have been
hypothesized to be  white dwarfs of mass $\sim$0.5
M$\odot$. The other survey group EROS however, mainly
from their non-detection (their initially reported events turned out to
be variable stars),
have recently claimed that this contribution is less than 20\% 
(Renault et al., 1996). 
The OGLE collaboration have recently extended their survey program and have 
begun a dedicated survey program towards both the LMC and the Galactic Bulge.
So more events will surely be detected by different groups and  
the situation will be clear as the spatial distribution and time scales of more
events are known. 

If indeed these events towards the LMC are due to stars, it opens a new 
possibility to look for planets around the LMC stars. In particular, since the
distance between the source and the lens in this case is smaller,
the Einstein ring radius R$_E$ of the star is smaller. 
Typically, in such a case, the Einstein ring radius R$_E$ can be written as

$$R_E = {D\over{100 pc}} \sqrt{M\over{M_\odot}} \ \ AU \eqno (1)$$

\noindent where D is the distance between the source and the lens, and 
M is the mass of the lens.

Thus the search for planets in such a case would be most sensitive to 
planets at a distance of about 1 AU from the star for a 1 solar mass lens.

The main obstacle to making a follow up monitoring program towards 
the LMC is that
the number of ongoing events at a given time is extremely small,
at the most 2 at present. 
Such a small  number of
events does not justify the dedicated allocation of a telescope 
for this program. If more events can be detected at a given time,
as indeed expected in the near future after the EROS II and OGLE 
survey programs towards the LMC have their alert systems fully operational, 
frequent monitoring of
ongoing LMC microlensing events is a promising possibility and follow 
up programs may soon be taken up.
This would make the search strategy sensitive to a very different region in the
orbital parameter space, and to a very different group of stars,
in this case being those within the LMC.

\subsection{Towards the Galactic Bulge}

The microlensing events towards the Galactic Bulge however tell a
different story. In this case, since the whole line-of-sight 
passes through the thick concentration of stars in the Galactic disk, 
the known stellar population contributes a great deal to the microlensing
optical depth. In fact, the original experiment suggested 
by Paczy\'nski (1991) was to check the experimental capabilities of the
proposed microlensing survey programs towards the LMC, by first
looking for such events towards the
Bulge where the known stellar density is bound to
cause microlensing. Such a test experiment was taken up 
by the OGLE group, who also reported their first discovery in 1992
(Udalski et al. 1993). More discoveries followed by the MACHO collaboration
(Alcock et al., 1995) and later by DUO  collaboration (Alard et al., 1995).  
Here again, there were surprises. The event rate in this case
was too high, and the derived optical depth came out to be larger than 
originally thought. It was soon realised that, the effect 
of Bulge stars acting as lenses was originally ignored, which partly 
explained the observed high optical depth (Kiraga and Paczy\'nski, 1994). 
Even after taking the Bulge-Bulge lensing into account, the event rate
was still too high. From the distribution of the events and from a mapping of the
microlensing optical depth, the presence of the Galactic bar was
rediscovered which, if inclined to the line of sight by about
15 degrees, could account for the
observed optical depth and the distribution of the events
(Paczy\'nski et al. 1994).

Thus, towards the Galactic Bulge at least, there is general consensus
on the fact that most of the lensing objects are stars, although more work 
is necessary to precisely determine what fraction of them belong 
to the Bulge and what fraction to the Galactic disk. The time scale of 
the events were used to model
the mass of the lensing stars (Zhao, Spergel and Rich, 1995) who found that
the lenses are consistent with being stars with mass larger than 
0.1 M$\odot$. 

\subsection{Confirmation of Stars as Lenses}

Out of the more than 100 microlensing events detected so far, only
8 are observed towards the LMC.  The rest overwhelming
majority are observed towards the Galactic Bulge for which the lenses are
believed to be due to stars in the line of sight. 

The fact that these are due to stars has been confirmed in every occasion
where the mass of the lensing star could be determined more accurately. 
There are at least 4 such examples which are the following.

\begin{enumerate}
\item  The binary event towards the LMC

Out of the 8 microlensing events so far observed towards the 
LMC, one is due to a binary lens. Analysis of this event conclusively 
proves that the lens in this case is a star within the LMC (Bennett et al.
1996).

\item  The parallax event towards the Galactic Bulge

A parallax event towards the Galactic Bulge was found, which made it 
possible to constrain the mass as well as the 
location of the lens. It was found to be a star
of mass in the range 0.4 to 2 M$\odot$ at a distance of 
1 to 4 kpc (Alcock et al. 1995).

\item  The binary events towards the Galactic Bulge

A few binary events have been found towards the Galactic Bulge,
namely OGLE \#7 (Udalski et al, 1994b), DUO \#2 (Alard et al., 1995b),
the data for which have been analyzed in detail. 
In each of these cases,
the mass of the lens is consistent with the lens being a
low-mass star.

\item Extended source towards the Galactic Bulge:

A giant star towards the Galactic Bulge was microlensed in 1996
(MACHO 95-30), which was spectroscopically monitored during the
event. The change in some spectral features, particularly the  H$\alpha$
line and the TiO bands,
provides conclusive evidence  that the lens is a low-mass star,
with its median mass being around 0.7 M$\odot$ (Alcock et al., 1997).

\end{enumerate}

Thus in each and every case where the lens mass could be
determined better than in a mere statistical sense, the lens has been 
found, without exception, to be a low-mass star. 
It is then a logical step to look for planets around these 
lensing stars through microlensing: the rest of this paper
deals with the details of such a method to search for extra-solar
planets.

\subsection{Theoretical Aspects of Stars as Lenses}

Before proceeding into the details of the lensing due to binaries and planets,
it is useful to review the basics of the lensing by a single star.  
For the details of the theoretical aspects of the lensing
by a star, the reader may refer to the excellent review
article by Paczy\'nski (1996) and the very exhaustive 
monograph devoted to the subject of Gravitational Lensing
by Schneider, Ehlers and Falco (1992). The basic information
which we will need later are essentially the following.

\begin{figure}
\centering \leavevmode
\epsfxsize=9truecm \epsfbox{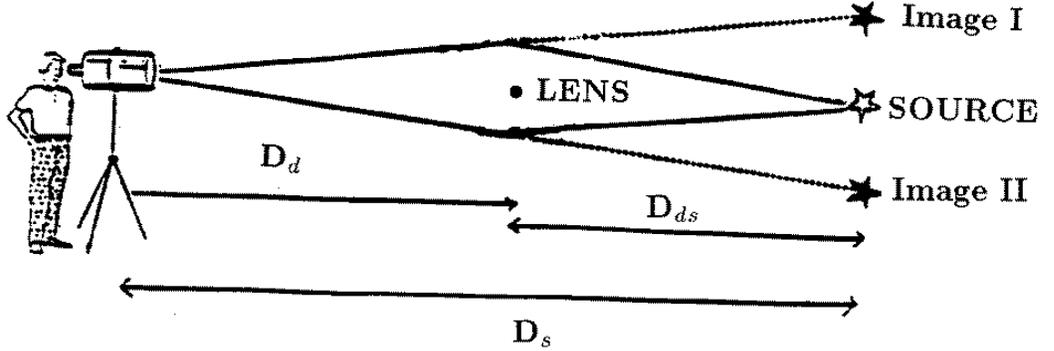}
\caption{Schematic geometry of the gravitational microlensing. 
The presence of the lens causes the image of the source to split into 
two, their combined brightness being always larger than that
of the unlensed image. Note that the deflection
due to the lens and the separation of the images are greatly
exaggerated in this schematic diagram.}
\end{figure}

With the lensing geometry as described in Figure 1, the
Einstein ring radius R$_E$ can be written as
$$ R_E^2 = {{4GMD}\over {c^2}}, D = {D_{ds} D_d\over D_s}  \eqno (2)$$
 
\noindent where M is the  the mass of the lensing object,\par
D$_d$ is the the distance to the lensing object, \par
D$_{ds}$ is the the
distance from the lens to the source, and\par
D$_s$ is the distance from the observer to the source.

The amplification due to the microlensing depends only on the
impact parameter, which can be written as 

$$A =  {{u^2+2}\over u(u^2+4)^{1/2}} \eqno (3)$$
where $u$ is the impact parameter in units of $R_E$.
 
This equation can be easily inverted to
derive the impact parameter from a given amplification
$$ u = 2^{1/2} [A (A^2 - 1)^{-1/2} -1]^{1/2} \eqno (4)$$
which can be used to derive the minimum impact parameter
$u_m$ from an observed light curve. 

\begin{figure}
\centering \leavevmode
\epsfxsize=8truecm \epsfbox{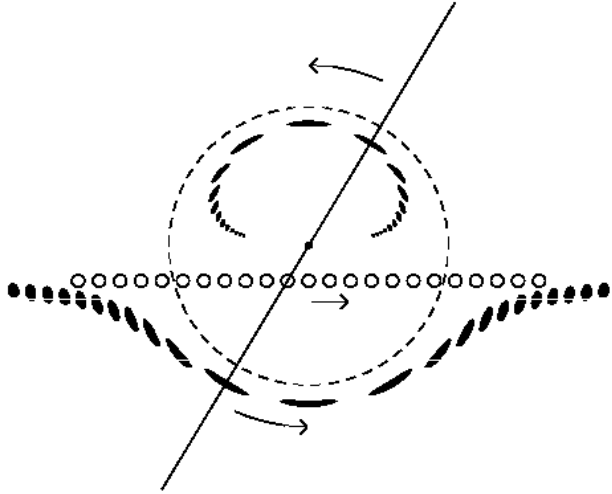}
\caption{ This figure shows how the
apparent positions and the sizes of the images change at various stages of the
microlensing. In this geometry the position of the lens, 
indicated by a solid dot,  is fixed, 
and the open circles show the actual positions of the source.
The filled circles show the images of the source as the source passes
close to the lens in the plane of the sky. The dashed circle is
the Einstein ring of the lens. 
At any instant, the source, the lens and the two images lie on a straight 
line.  (Taken from Paczy\'nski, 1996)}
\end{figure}

The the time scale of microlensing is the time taken
by the source to cross the Einstein ring radius, which 
is given by
  
$$ t_0 = {R_E \over{V_e}}  \eqno (5)$$
where $V_e$ is the tangential velocity of the lensing object.
The impact parameter
at any time during the microlensing event can be expressed as
$$ u = [u_m^2 + ({t-t_{m}\over t_0})^{2}]^{1/2} \eqno (6)$$
where $t_{m}$ is the time corresponding to  the minimum impact parameter 
 (or the maximum amplification).



From Eq. 2 and 5, the mass of the lens can be expressed as
$$M = {[t V_e c]^2 \over {4GD}} \eqno (7)$$

\begin{figure}
\centering \leavevmode
\epsfxsize=8truecm \epsfbox{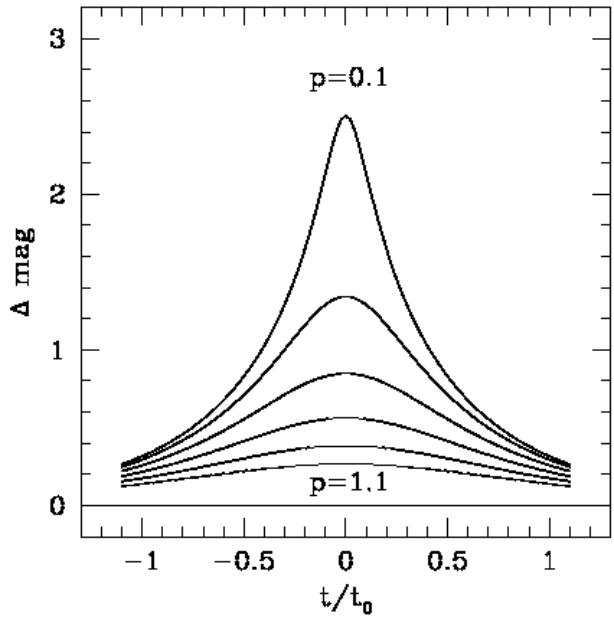}
\caption{The microlensing light curves as a function of impact parameter. }
\end{figure}

\subsection{Effect of Extended Source}

In case of the microlensing events towards the LMC and the Galactic Bulge,
the point-source approximation may not always be valid. This is particularly 
the case if the LMC events are caused by the LMC stars and the Bulge events 
are caused by the Bulge stars, in which case the distance between the source 
and the the lens is not large. Consequently, the Enstin ring radius is smaller,
in which case the source size cannot be neglected. (For more details 
see Sahu, 1994b). This is also very important for lensing caused by planetary 
mass objects since the Einstein ring radius of a planet may not always be 
much larger than the size of the source. 
In such a case, different parts of the source will be amplified differently and the net amplification can be expressed as (Eq. 6.81 of Schneider,
Ehlers and Falco, 1992)
$${\int d^2y \ I(y) \ \mu_p(y) \over \int d^2y \ I(y)} \eqno (8)$$
\noindent where I(y) is the surface brightness profile of the source,
$\mu_p(y)$ is the amplification of a point source at point
$y$, and the integration is carried out over the entire surface
of the source.

In extended-source approximation, since different parts of the source 
are amplified differently, the limb darkening effect can be important. 
This can make the event chromatic and the ratios of the emission/absorption
in the star features in the source star can vary during the event (Loeb and Sasselov, 1995). Such effects have 
indeed been seen in case of MACHO 95-30 (Alcock et al. 1997).
The extended source effect can be particularly important
in case of planetary events where, in general, the source-size
cannot be neglected.  

If the source can be approximated as a disk of uniform brightness, then the
maximum amplification, when the source and the lens are perfectly aligned, is given by

$$ A_{max} = [1 + {4R_E^2\over{r_{0}^2}}]^{1\over{2}} \eqno (9)$$
\noindent where $r_0$ is the radius of the source. When the Einstein
ring radius is the same as the radius of the source, the maximum 
possible amplification in such a case is $\sim$2.24.

\section{PLANETS AS LENSES}

\subsection{Observational Characteristics}

The light curve due to a binary lens, unlike the single lens, can be complex 
and can be very different from the mere superposition
of two point lens light curves.
In case of a double lens, the lens equation, which is 
a second order equation for a single lens, becomes
two 5th order equations (or one 5th order equation in the complex plane,
Witt and Mao, 1995).  
The most important new feature is the formation of 
caustics, where the amplification is infinite for a point source,
but finite for a finite size source. When the source crosses
a caustic, an extra pair of images forms or disappears. 
A full description of the microlensing due to a 
double lens is given by Schneider and Weiss (1986).

\begin{figure}
\centering \leavevmode
\epsfxsize=8truecm \epsfbox{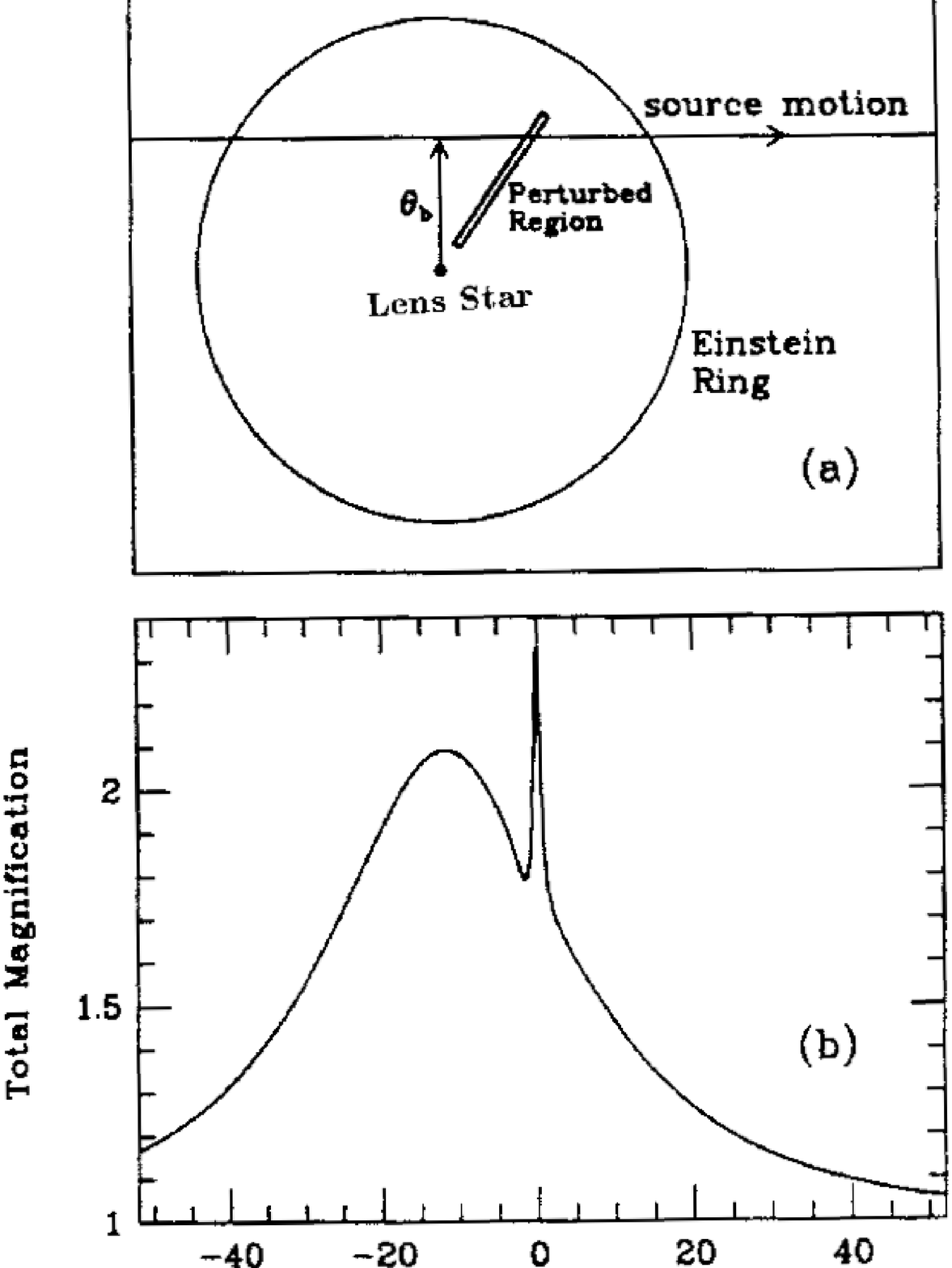}
\vspace*{0.5cm}
\caption{ Fig. (a) at the top schematically shows the geometry of the 
star-plus-planet lensing event
and fig. (b)  shows the resulting light curve. 
The mass of the planet is 10$^{-3}$ times that of the primary,
and is situated close to the Einstien ring of the primary
(adapted from Gould and Loeb, 1992).}
\end{figure}

If the lensing star has a planetary 
system, the effect of the planet on the microlensing 
light curve can be treated as that of
a binary lens system. The signature of the planet can be seen,
in most cases, as sharp extra
peaks in the microlensing light curve. Computer codes for analysis of
such data have been developed by Mao and Di Stifano (1995)
and Dominik (1996).

\begin{figure}
\centering \leavevmode
\epsfxsize=11truecm \epsfbox{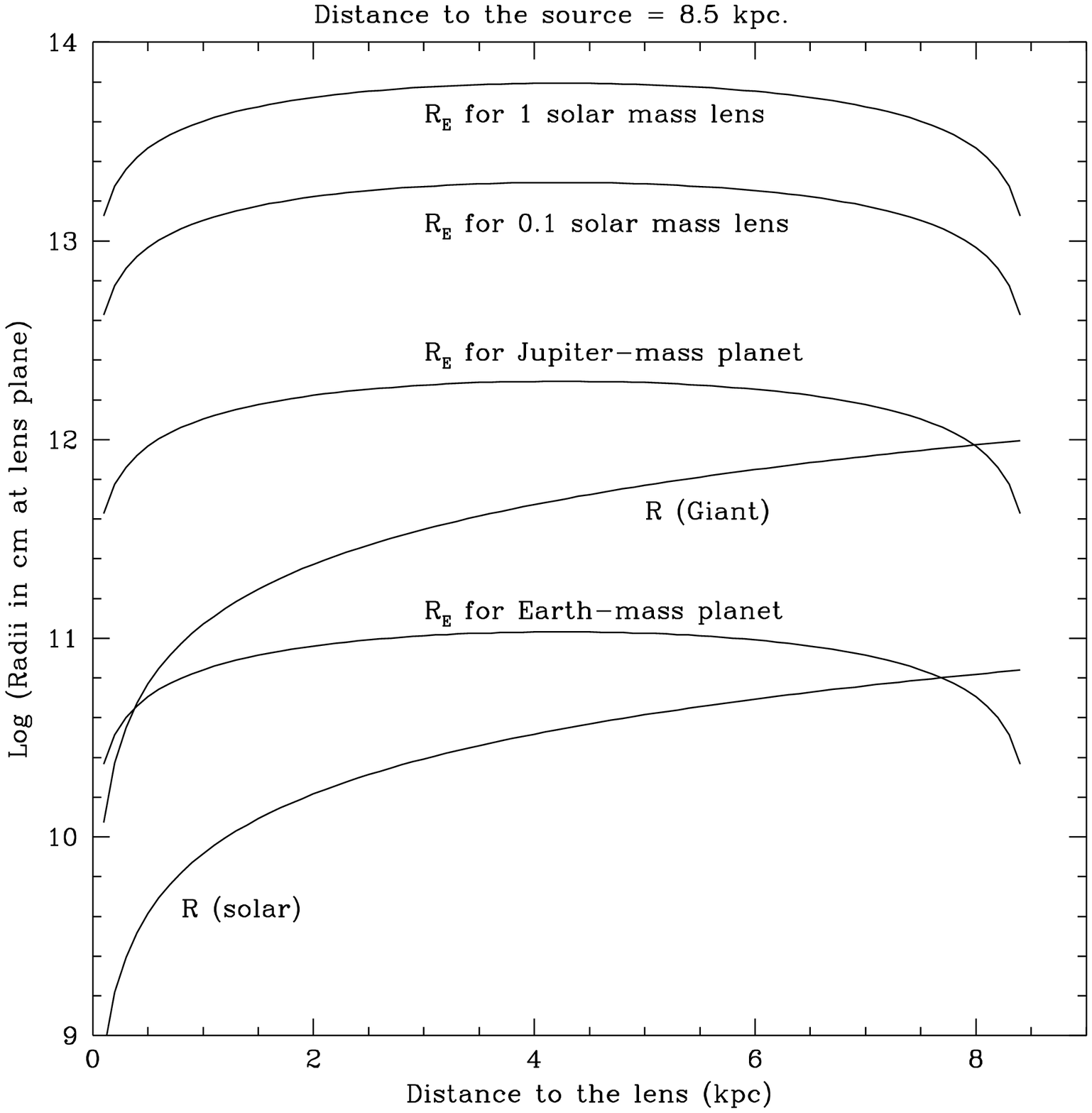}
\caption{The figure shows the sizes of the Einstein ring radii R$_E$ at the lens plane,
for a lensing event towards the Galactic Bulge. R$_E$
for an Earth-mass planet a solar mass star are shown.
Also shown are the actual radii of a solar type star and a typical giant star
as projected onto the lens plane, which are denoted by R(solar) and R(giant)
respectively. For a Jupiter-mass planet, R$_E$ is
almost always larger than the radius of a giant star, so the effect 
of Jupiter can always be significant.  But for an Earth-mass planet, 
there is only a small 
parameter space where the Einstein ring radius is larger than the size
of a giant star. 
If the source is  a main-sequence star like our Sun, the
Einstein ring radius due to an earth-mass planet is almost always larger 
than the source size, and hence the amplification can be large.}
\end{figure}

\begin{figure}
\centering \leavevmode
\epsfxsize=11truecm \epsfbox{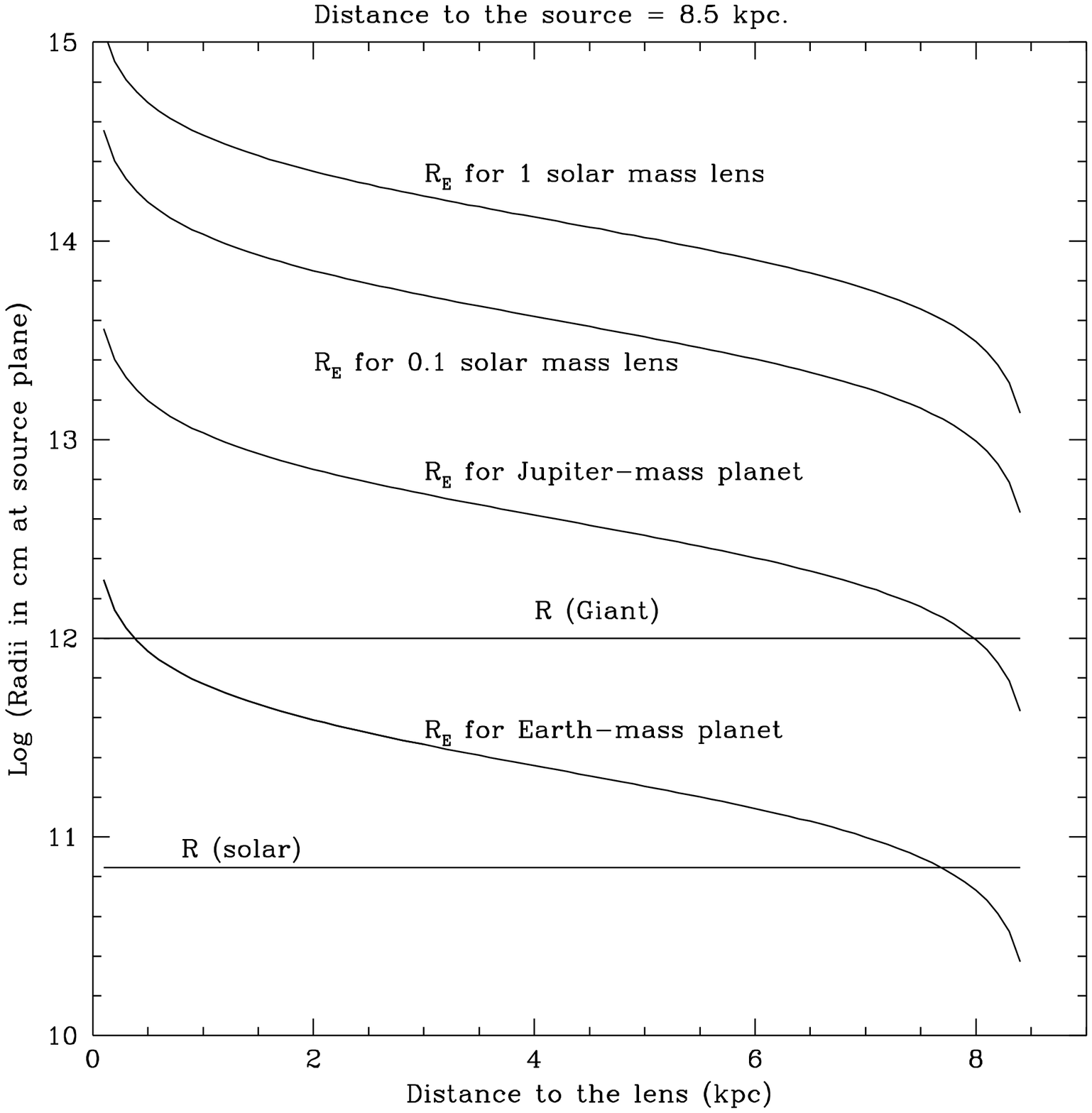}
\caption{ The same as Fig. 5, but here the size of the Einstein ring 
radii and the sizes of the sources are all as seen projected onto the 
source plane.}
\end{figure}

The effect of a double star or a planetary system on the microlensing light
curve was first investigated by
Mao and Paczy\'nski (1991). They showed that about ~10\% of the 
lensing events should show the binary nature of the lens,
and this effect is strong even if the companion is a planet.
 
The problem of microlensing by a star with a planetary system 
towards the Galactic Bulge was further
investigated by Gould and Loeb (1992). They noted that, for a 
solar-like system  half way 
between us and the Galactic Bulge, Jupiter's orbital radius coincides
with the Einstein ring radius of a solar-mass star.
Such a case is termed `resonant lensing' which increases the
probability of detecting the planetary signal. In $\sim$20\% of the cases, 
there would be a signature with magnification larger than 5\%.
 
The importance of  the resonant lensing can be qualitatively 
understood as follows. 
In Fig. 2, the impact parameter changes through a large range as the
source passes close to the lens. The positions of two images
formed by the lensing effect change continuously, but they remain close 
to the Einstein ring for a large range of impact parameters. 
So, the effect of the planet can be large if the planet happens to be 
close to the Einstein ring, which causes a further amplification. This also
qualitatively explains why the probability of observing the
effect of the planet increases if it is close to the 
Einstien ring.
 
In a large number of cases however, the resulting light curve
due to a planet plus star system is close to the superposition of
two point lens light curves (Fig. 4). This is particularly true
when the star-planet distance is much larger than R$_E$.
In such a case, the time scale 
of the extra peak due to the planet, $t_p$, and the time scale of the
primary peak due to the star, $t_s$, are related through the relation
$t_p / t_s = \sqrt{(m_p/M_s)}$ where $m_p$ is the mass of the planet and $M_s$
is the mass of the star. Furthermore, if the source size cannot be
neglected, the maximum amplification
given by Eq. 9 remains valid. Figs. 5 and 6 show the sizes of the
Einstein ring radii due to planetary and stellar mass lenses as a function of
distance to the lens. 
The typical sizes of the main-sequence and giant sources
are also shown. In such a case, it is clear that  
the size of the source is almost always smaller than the Einstein
ring radius of a Jupiter-mass planet, as a result
the amplification due to the planet can be large.
The amplification can also be
large for an Earth-mass planet if the source is a main sequence star.
However, if the source is a giant-type star, then there
is only a fixed range of D$_d$ where the amplification due to
an Earth size planet can be large.

In general. the situation is different in case of formation of caustics.
Fig. 7 shows the effect of the formation of caustics
and consequent high amplification and sharp peaks
caused by the planets. The planets, in this case, are
situated at different orbital radii and the mass of each planet 
is 10$^{-3}$ times that of the primary. 
The solid curve shows the light curve without the presence of the
planets. The dashed and the dotted curves correspond to two 
representative tracks of the source {\it with} the presence of the
planets (Wambsganss, 1997). 

\begin{figure}
\centering \leavevmode
\epsfxsize=12truecm \epsfbox{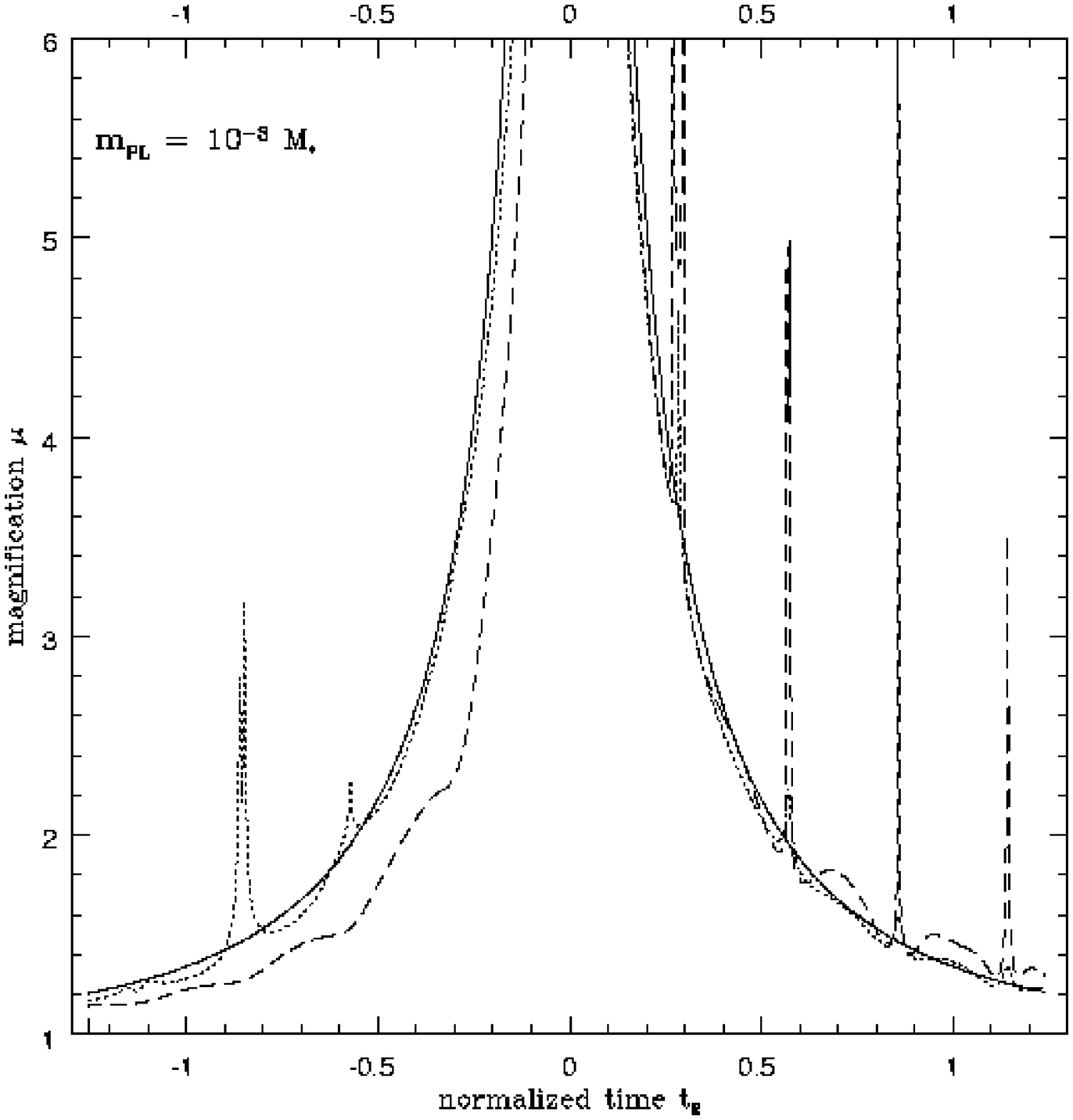}
\caption{Effect of a few planets situated at different orbital radii on the
microlensing light curve. The mass of each planet is 10$^{-3}$ times that 
of the primary. The projected distances of the planets from the primary,
in units of R$_E$, are 0.57, 0.65, 0.74, 0.86, 1.16, 1.34, 1.55 and 1.77.
The solid curve shows the light curve without the presence of the
planets. The dashed and the dotted curves correspond to two 
representative tracks of the source {\it with} the presence of the
planets. This shows how the influence of the planet on the
microlensing light curve can be both upward and downward,
and illustrates the possible high amplification and sharpness of the
extra peaks caused by the planets (taken from Wambsganss, 1997).}
\end{figure}

The minimum duration of the extra feature due to the planet, to a first 
approximation, is the time taken by the source to 
cross the caustic, which can be about 1.5 to 5 hrs. The maximum duration 
of the spike is roughly the time taken by the planet to cross its own Einstein ring.
Using a reasonable set of parameters (the lower mass of the planet is
taken as that of the Earth, the higher mass is assumed to be that of Jupiter)
this can be a few hours to about 3 days. 
 Any followup program must be accordingly adjusted so that the extra
feature due to the planet is well sampled.

\subsection{Theoretical Work}

A full description of the theoretical aspects of planets acting as lenses
is beyond the scope of this review. 
To date, there are a few countable number of papers which deals with the 
theoretical prediction of planetary signals on the light curve,
which the reader may refer to (Bolatto and Falco, 1995; Bennett and 
Rhie, 1996; Wambsganss, 1996; and Peale, 1996, this volume). 

\subsection{Requirements for a Follow-up Network}

The first requirement for a followup network is access to the `alert' 
events. With the alert capability of the survey programs firmly in place, 
it is now possible
to build a follow-up network. At present, the alert events from the MACHO collaboration at a given time is sufficient to carry out a ground based 
follow up program with small telescopes towards the Galactic Bulge. 
After OGLE and EROS II experiments have their alert systems in place, 
the number of on-going alert events at a given time will increase and
it may be possible to extend such followup networks to larger telescopes, and
also perhaps towards the LMC.

The second requirement is the ability to monitor hourly. It should be 
noted that, assuming that the longer time scale events are mostly due to slower 
proper motion of the lensing star, the time scale of the planetary signal 
approximately scales with the time scale of the main event. So the monitoring, 
in general, can  be less frequent for longer time scale
events.  But typically, as noted before, the time scales of the planetary event 
can be a few hours to a few days. So the followup monitoring 
program must have the capability to do hourly monitoring 
so that the extra feature due to the planet
is well sampled. For discrimination against any other short term variations,
some color information is also useful, since the microlensing is expected 
to be achromatic, where as most other types of variations 
are expected to have some chromaticity. Thus it is preferable to have 
a few observations in two colors. 

The third requirement is to have 24-hour coverage in 
the monitoring program. This calls for telescopes at appropriately spaced 
longitudes around the globe.

\subsection{PLANET Collaboration}

PLANET (Probing Lensing Anomalies NETwork) is the first such collaboration, 
which was formed in 1995, soon after the alert capability was in place. 
PLANET currently uses 4 telescopes situated at appropriately spaced 
longitudes around the globe in order to achieve 24-hour coverage in the 
monitoring program. PLANET has now completed 3 years of observing 
campaigns. More details of the 
present status of the PLANET collaboration can be found in Albrow et al. in
this volume, which also lists the members of this collaboration
in alphabetical order as the author-list.
\vskip 0.2cm
The present capability of the PLANET collaboration is the following.

\begin{enumerate}
\item  Photometric accuracy of less than 5\% is routinely observed.

\item Hourly monitoring, and (weather permitting) close to 24-hour
coverage in the monitoring program is achieved.

\item Online reduction facilities have been developed. 

\end{enumerate}

We hope to be able
to provide `secondary alerts' in the near future. This may allow the 
interesting events to be more intensely followed, both photometrically and 
spectroscopically, by other observers.

A binary event was  seen from the PLANET data, with a photometric accuracy of 
better than 5\%  both in V and I bands. Furthermore, many new short term 
variables have been  seen from the data.

We hope to be able to provide `secondary alerts' in the near future. 
This may allow the interesting events to be more intensely followed, 
both photometrically and spectroscopically, by other observers.

The information on the PLANET network can be found at the WWWeb sites:

http://www.stsci.edu/$\sim$ksahu/PLANET.html

http://www.astro.rug.nl/$\sim$planet

\subsection{Other Followup Collaborations}

GMAN is another recent collaboration whose aim is to do similar
followup work using ground based telescopes
(Pratt et al. 1996).   A bigger scale, NASA-sponsored monitoring program 
is also planned by Tytler et al. 

In future, it may also be possible to install a robotic telescope 
at the south pole,
which will be ideal to carry out such a followup observational program 
in the Bulge season. The Bulge season conveniently falls 
at the time of the southern winter so that the telescope can be
used 24-hours a day for the monitoring program.
This would avoid the necessity of many different telescopes at
different longitudes, although it presents its own
obvious problems of logistics which have to be overcome.

\subsection{A Cautionary Tale}
I started this review with a laudable remark on the gravitational effect 
by saying that the gravitational effect, unlike the other effects,  
makes use of the nearby star, thereby making the search 
enormously easier and this has been the cause for most of the other earlier 
discoveries of planets.
I would like to end this review with a cautionary tale, 
a part of which has been 
told earlier in some other context, but is very relevant here.

In spite of all the simplicity of the gravitational effect, it can, and 
indeed had, its surprises. The perturbation in the orbit of Uranus
was simple to interpret as due to an unseen planet which led to the 
discovery of Neptune. But the precession in the orbit of Mercury was 
also first interpreted 
as due to an unseen planet, which was named `volcan'. 
All attempts to search for the illusive `volcan' however failed, 
and it took Einstein's general theory of relativity to correctly understand the 
the precession of Mercury. 

Another, perhaps less elegant, story was repeated in 1992, after
the first `planet' was discovered around the pulsar PSR1829-10  from a 
periodic variation in the pulsar period (Bailes et al.
1991). This variation was however later found to be due to an
error in the algorithm used, which did not take the 
orbital motion of the Earth correctly into account (Lyne and Bailes, 1992).

At the time of this writing, similar claims have been made on the presence of 
a planet around 51 Peg: line width variations in a line originating in the 
atmosphere of 51 Peg have been seen, the periodicity of is consistent with the 
orbital period of the putative planet. Based on this observation, it has 
been claimed that the observed radial velocity variation in 51 Peg could be
due to non-radial pulsations rather than due to a planet (Gray, 1997).

The point here is not to make any judgement on the presence or absence 
of a planet around 51 peg, which undoubtedly will be resolved soon with 
more data, but the point is to emphasize the fact that the interpretation 
of this apparently simple effect may not always be easy. Similar is the 
case with the gravitational microlensing effect due to a planet. 
The planetary signature
may not always be a unique solution, some photometric deviations in one
or two out of millions of data points may be unavoidable,
some double source or blending effects may
be mistaken to be due to a binary lens: the list is long and
and any of these 
may some times be misinterpreted as due to planets.
All such effects must be properly taken into account  
for a correct interpretation, before any claims are made
of the presence of planetary  companions around lensing stars.

\acknowledgments

I would like to thank all the members of the PLANET collaboration, for their
help and suggestions.

\end{document}